# Flux-pinning mediated superconducting diode effect in the NbSe$_2$/CrGeTe$_3$ heterostructure

A. Mehrnejat[1†], M. Ciomaga Hatnean[2], M. C. Rosamond[3], N. Banerjee[1,4], G. Balakrishnan[2], S. Saveliev[1], and F. K. Dejene[1†]

[1] Department of Physics, Loughborough University, Loughborough, LE11 3TU, UK.

[2] Department of Physics, University of Warwick, Coventry, CV4 7AL, UK.

[3] School of Electronic and Electrical Engineering, University of Leeds, Leeds, LS2 9JT, UK.

[4] Blackett Laboratory, Imperial College London, London, SW7 2AZ, UK.

†Corresponding authors: a.mehrnejat@lboro.ac.uk and f.dejene@lboro.ac.uk

## Abstract

In ferromagnet/superconductor bilayer systems, dipolar fields from the ferromagnet can create asymmetric energy barriers for the formation and dynamics of vortices through flux pinning. Conversely, the flux emanating from vortices can pin the domain walls of the ferromagnet, thereby creating asymmetric critical currents. Here, we report the observation of a superconducting diode effect in a NbSe$_2$/CrGeTe$_3$ van der Waals heterostructure in which the magnetic domains of CrGeTe$_3$ control the Abrikosov vortex dynamics in NbSe$_2$. In addition to extrinsic vortex pinning mechanisms at the edges of NbSe$_2$, flux-pinning-induced bulk pinning of vortices can alter the critical current. This asymmetry can thus be explained by considering the combined effect of this bulk pinning mechanism along with the vortex tilting induced by the Lorentz force from the transport current in the NbSe$_2$/CrGeTe$_3$ heterostructure. We also provide evidence of critical current modulation by flux pinning depending on the history of the field setting procedure. Our results suggest a method of controlling the efficiency of the superconducting diode effect in magnetically coupled van der Waals superconductors, where dipolar fields generated by the magnetic layer can be used to modulate the dynamics of the superconducting vortices in the superconductors.

## Main text

Type II superconductors (type II SCs) are distinguished by their characteristic mixed state in the phase diagram due to the appearance of superconducting vortices[1]. The dynamics of these Abrikosov vortices can be modified by applied current, external magnetic field, sample geometry/confinement[2,3], and magnetic proximity effect[4–6], thereby opening various possibilities for controlling device properties, for instance, the superconducting diode effect (SDE)[7–9]. The appearance of nonreciprocal current-voltage characteristics in a superconductor has received revived interest following the observation of near-field-free superconducting rectification effects in van der Waals heterostructures (vdWHs) and multilayered Rashba superconductors[10,11]. These SDEs were reported to arise from breaking inversion symmetry (IS) and time-reversal symmetry (TRS)[12]. Asymmetric critical currents have also been observed in other works using conformally mapped nanoholes[13], size confinement in constrictions[2], and Josephson junctions[14,15]. It is plausible that asymmetric *I-V* characteristics can arise from extrinsic properties such as interface or edge roughness in lithographically defined constrictions[16] or other intrinsic mechanisms such as asymmetric vortex-flow energy barriers[1,17,18], magnetic flux penetration[20,21], flux pinning effects[19,22], vortex limited critical currents[23] and avalanches[24,25]. Hence, identifying the key driving mechanisms of the SDE and establishing their relations to other nonreciprocal transport effects is of paramount importance for future superconducting electronic as well as spintronic devices. Atomically flat interfaces in artificially engineered vdWHs, which have been integral in the discovery of various emergent quantum effects in spintronic, thermoelectric, superconducting, and optical applications[26–28], could offer unique possibilities for a more detailed understanding of the mechanisms of SDEs by excluding extrinsic interfacial mixing or disorder.

In this work, we demonstrate the observation of SDE in the vdWH of $NbSe_2$/$CrGeTe_3$(CGT), where the magnetic domain walls in CGT offer a unique possibility for guiding superconducting vortices[29,30] essential for magnetically controlled pinning of vortices at the $NbSe_2$/CGT interface[19]. $NbSe_2$ is an anisotropic layered superconductor extensively studied for its unique superconducting and spintronic properties in thicknesses down to the monolayer regime[2,31–35]. CGT is a 2D magnetic semiconductor ($E_g \sim 0.38$ eV)[36] with a Curie temperature of approximately 65 K and exhibits diverse thickness-dependent magnetic domain transformations[37,38]. The magnetic coupling of superconducting $NbSe_2$ with CGT allows us to exploit its topological vortex phase transitions by modifying the confinement potentials[39] of Abrikosov vortices and, in turn, controlling the SDE via asymmetric flux-pinning at the $NbSe_2$/CGT and $NbSe_2$/$SiO_2$ interfaces[23].

Figure 1(a) shows a schematic diagram of our $NbSe_2$/CGT vdWH device along with the electrical configuration for longitudinal resistance measurement. The $NbSe_2$ crystals (≥99.999% purity) were obtained from Ossila Ltd., and the CGT (≥99.99% purity) crystals were grown using the flux method as described in Ref.[40]. Figure 1(b) shows an example micrograph of one of the measured devices (named device B) comprising two adjacent $NbSe_2$/CGT and bare $NbSe_2$ devices, which we measured at the same time under identical experimental conditions in a Cryogenic Ltd. cryostat. We adopted a top-down device fabrication approach where the vdWH ($NbSe_2$/CGT) was transferred onto a prepatterned Si/$SiO_2$ substrate with Ti/Au (45 nm) electrical contacts patterned using standard UV lithography and e-beam evaporation. In the first step of our device fabrication, $NbSe_2$ and CGT crystals were mechanically exfoliated onto two different polydimethylsiloxane stamps (from Teltec Ltd.) under ambient conditions. After identifying homogenous flakes, we first transferred the $NbSe_2$ flake on the contacts, followed by the CGT flake transfer on one side of the NbSe$_2$ flake. We fabricated and studied several devices with NbSe$_2$ thicknesses ranging from 5 nm up to 45 nm, while device degradation due to the oxidation of flakes was minimized by completing flake-transfer processes in under an hour.

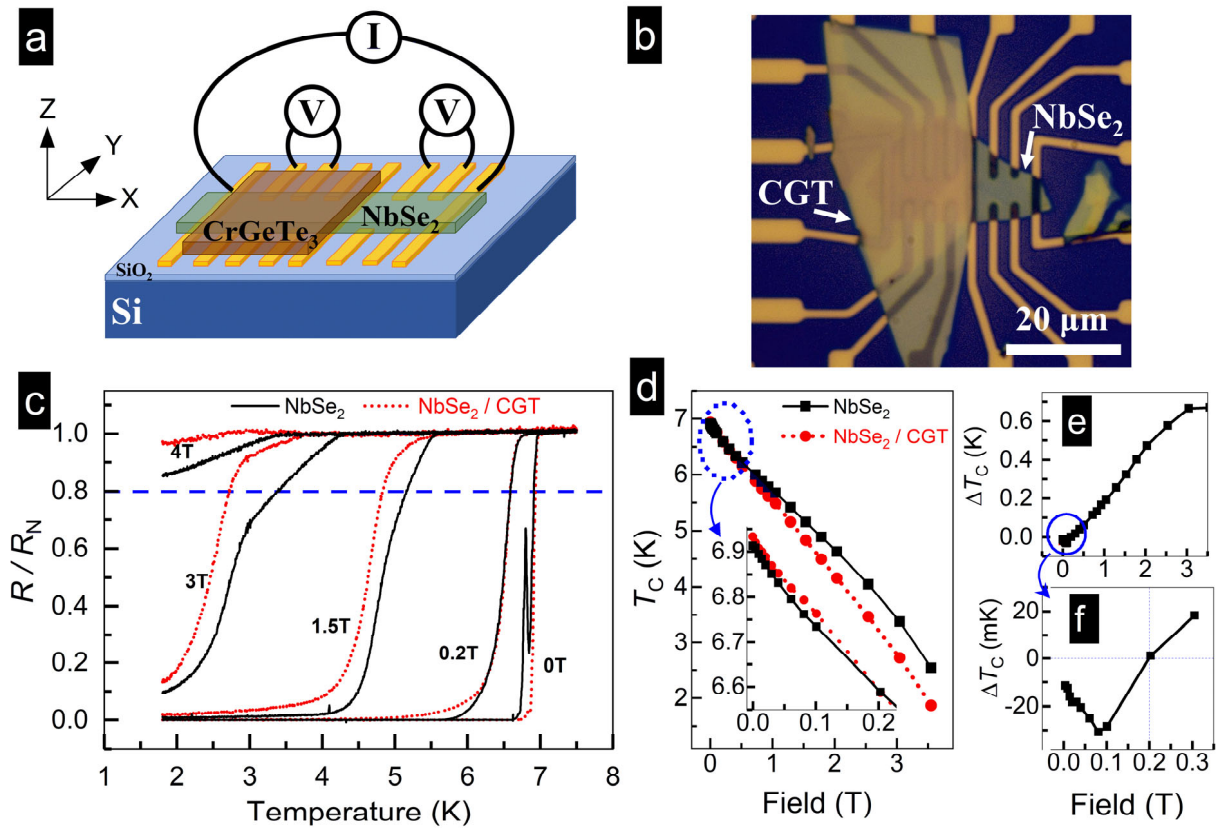

Figure 1. a) Schematic of the measured NbSe$_2$ flake on two adjacent Hall bar electrodes. A DC current was sourced in two outer electrodes, while two nanovoltmeters were used to simultaneously measure the bare NbSe$_2$ and CGT-capped NbSe$_2$ sections. b) An optical micrograph of a device. c) Temperature-dependent resistance of bare NbSe$_2$ (solid black) and NbSe$_2$/CGT (dashed red) for different applied out-of-plane magnetic field values. d) Field dependence of the critical temperature ($T_C$) in bare NbSe$_2$ and NbSe$_2$/CGT at an applied current of 50 µA. The $T_C$ values were taken at $R/R_N$ = 0.8. The inset shows a magnified plot of the circled low-field range. The $T_C$ of the bare NbSe$_2$ is smaller in this magnetic field range, but this changes after the CGT magnetization saturates at approximately 200 mT. e) The difference between bare and capped NbSe$_2$ critical temperatures $\Delta T_c$ versus B. f) Low-field plot of data in (e) that peaks around the field range 80-100 mT followed by a sign change.

We performed temperature-dependent resistance measurements (RT) of both the bare and capped NbSe$_2$ sides. The current was applied along the x-direction (as shown in Figure 1(a)) parallel to the ab-plane of NbSe$_2$, while an out-of-plane external magnetic field was applied along the z-direction parallel to the c-axis of the flake. Figure 1(c) shows the RT of a device (labeled device A) for various applied magnetic fields. The zero-field critical temperature of the bare NbSe$_2$ ($T_C$~6.93 K) is just slightly larger than the CGT-capped NbSe$_2$ critical temperature ($T_C$~6.91 K); otherwise, both $T_C$ values are close to that of bulk NbSe$_2$, indicating the high quality of our devices, as confirmed by the residual resistivity ratios [*RRR* = *R*(room temperature)/*R*(above *T*$_c$)] ranging from 10 to 20, typical for exfoliated NbSe$_2$ flakes[32,41]. Figure 1(d) depicts the field dependence of the *T*$_C$ extracted based on the 80% line of the normalized resistance plots in Figure 1(c). Here, we can observe that for B<200 mT, the critical temperature of NbSe$_2$/CGT is slightly higher than that of bare NbSe$_2$ (see the inset in Figure 1(d)). On the other hand, when B>200 mT and CGT is saturated, the *T*$_C$ of bare NbSe$_2$ is larger than that of NbSe$_2$/CGT. This corresponds to the magnetization saturation field of the bulk CGT crystal (later shown in Figure 4(c)). For the bare side, in the low-field range, the transition is accompanied by the appearance of intermediate kinks previously attributed to phase slips or inhomogeneous strain distribution[42,43]. To illustrate this further, we plotted $\Delta T_C = T_C^{\text{NbSe}_2} - T_C^{\text{NbSe}_2/\text{CGT}}$ in Figure 1(e) and 1(f), where a clear sign change of $\Delta T_C$ at B=200 mT corresponds to the magnetization saturation point of CGT. It is also worth noting that for $\Delta T_C < 0$ values in Figure 1(f), the minimum is located in the field range of 80-100 mT, which, as we show later, plays a major role in the modulation of the critical current via the magnetic domain reorientation of CGT or flux-pinning[38,44].

Next, we present magnetoresistance (MR) measurements of our devices with an out-of-plane (OOP) external magnetic field (see Figure 2(a)). The CGT/NbSe$_2$ MR curve exhibits hysteresis with different *H*$_C$ values for trace and retrace measurements. This hysteretic MR feature, which is completely absent in the bare NbSe$_2$ flakes, can be attributed to flux pinning in NbSe$_2$ by the CGT domains[19] and corresponds to the field regime where a sign change in $\Delta T_C$ was observed in Figure 1(e) and (f). To identify the asymmetry in the trace and retrace *H*$_C$ values, we define $H_C^{\text{tr1}}$ ($H_C^{\text{tr2}}$) as the positive (negative) *H*$_C$ value when sweeping the field from 0.5T to -0.5T, as shown in Figure 2(a), where the solid curve represents the trace measurements. Similar parameters are defined on the retrace measurement (dashed curve) from -0.5T to 0.5T. When tracing/retracing the field from higher magnitudes toward 0T, there is a sharp transition from the mixed state to the Meissner phase; however, the transition is smoother when increasing the field value from 0T to higher magnitudes, indicative of stronger vortex pinning. To observe how these lower critical fields develop by increasing the applied current, we performed MR measurements for varying currents. The current dependence of $|H_C^{\text{tr2}}|$ and $|H_C^{\text{retr2}}|$, obtained from the extrema of the first field derivative of MR and plotted in Figure 2(b), does not show an appreciable difference over the measured field range. Nevertheless, based on this observation and by comparing several *I-V* measurements collected in the flux-pinning window, we can estimate the magnitude of the pinned flux in NbSe$_2$. We consider a series of *I-V* measurements collected at 100 mT while the field is set to this value through two different scenarios: first, by coming from the higher field of +150 mT (solid black line in Figure 2(c) noted as *V*$_{\text{HL}}$), and second, by coming from the lower field of 0T (dashed red line, *V*$_{\text{LH}}$ for lower to higher). As shown in the inset here (adapted from Figure 2(a)), one of these two *I-V* sweeps represents the Meissner phase, and the other is performed in the dissipative phase of the Abrikosov lattice inside the hysteretic flux-pinning regime. The observed opening between the two *I-V* curves arises at currents above 0.5 mA, which also agrees with the appearance of hysteresis in the MR measurements. We can estimate the amount of this trapped flux to be approximately 20 mT by observing the relative overlap of the *I-V* curve at an applied field of 120 mT, which was set to this value by coming from an initial 0T (blue dashed line in Figure 2(c)). Figure 2(d) depicts the difference between $V_{HL}$ and $V_{LH}$, which shows how this difference changes the sign for field values inside and outside of this pinning window resulting from the modulation of the critical current (the inset shows the MR for another device).

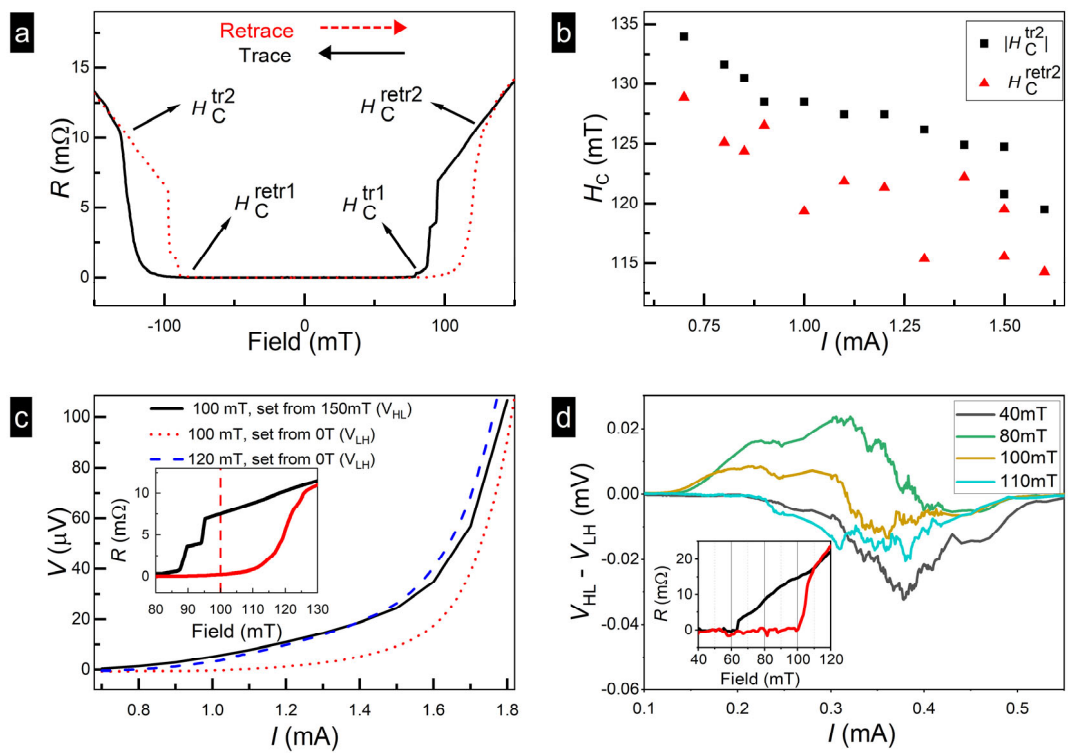

*Figure 2. a) Magnetoresistance measurements of NbSe$_2$/CGT showing hysteresis at 1 mA and 2 K. For traces (black) and retraces (dotted red), the definition of lower critical fields is indicated by arrows. b) Comparison of $H_C^{retr2}$ and $|H_C^{tr2}|$ values as a function of the applied current showing a relatively identical trend. c) Magnetic-field history dependence of the I-V curves at 100 mT (see inset) obtained by following the two field-setting scenarios $V_{HL}$ (higher to lower field, solid black) and $V_{LH}$ (lower to higher field, dotted red). The large dashed blue curve is the I-V data for an applied field of 120 mT ($V_{LH}$). The three I-V curves can be used to estimate a trapped flux of ~20 mT. d) The difference $V_{HL}$-$V_{LH}$ between measured I-V curves for similar field-setting scenarios shown in (c) highlights the role of flux-pinning for 80 mT and 100 mT, as shown in the inset MR.*

After establishing the impact of the CGT magnetization on the superconductivity of NbSe$_2$, we next discuss the characterizations of the SDE in the NbSe$_2$/CGT heterostructure side of device B. This device comprises a thin NbSe$_2$ flake (5 nm) and exhibits smaller critical currents, as expected for thinner superconductors[45]. As such, we limited the maximum applied current to below 2 mA to maintain high device quality and temperature-dependent stability. Figure 3(a) illustrates a typical trace and retrace *I-V* characterization of NbSe$_2$/CGT at an applied out-of-plane field of 20 mT. The trace curves (solid black) were recorded by sweeping current first from zero to the maximum positive current and second from zero to the maximum magnitude of the negative current. The retrace measurements (dashed red) from the maximum magnitudes back to 0 are not the focus of this paper but represent the retrapping current[9,46]. The positive critical currents (here: $I_C^+$ = 1208±1 µA) and negative critical currents ($I_C^-$ = -1200±1 µA) on positive and negative sweeps are different, indicating nonreciprocal transport of this SDE device with a nonreciprocity window of nearly 8 µA. Figure 3(b) depicts this window more clearly through the reflection (through the origin) of the negative critical current to the first quadrant. As mentioned above, these two curves are derived from the trace measurements where the current has been swept from lower to higher current magnitudes; thus, heat dissipation processes are relatively identical. As a proof of concept for SDE, we show the device rectification measurements in Figure 3(c), in which the input signal is a current pulse with symmetric positive and negative values of approximately $\pm$1.2 mA that fit in the nonreciprocity window shown in Figure 3(b), and the measured output voltage plotted with red dashed lines clearly demonstrates the rectifying behavior.

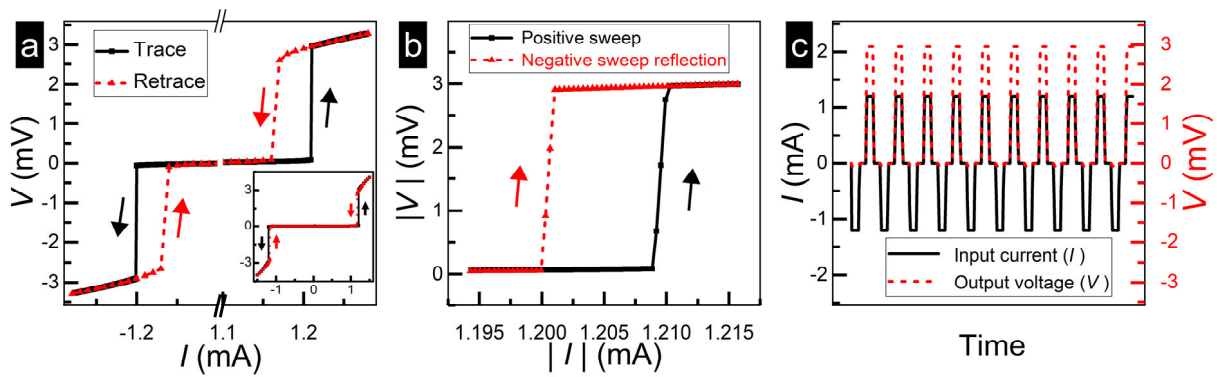

*Figure 3. Superconducting diode effect. a) An I-V characterization at B=20 mT and T=2 K. All measurements were carried out initially from zero to maximum current magnitudes (black solid trace curves) and then back to zero (red dashed retrace curves). The inset shows the full range of this I-V curve up to ±1.5 mA. b) Negative current sweep (trace) from the third quadrant of (a) has been reflected through the origin to the first quadrant and plotted along with the positive sweep, indicating clear asymmetry. c) Demonstration of the rectification effect in an SDE device for an input current signal (left axis). The device remains superconducting for -1.2 mA but turns normal for +1.2 mA, as shown using the corresponding output voltage (right axis).*

Figure 4(a) shows the dependence of the positive critical currents ($I_C^+$) and the absolute value of the negative critical currents ($I_C^-$) on the applied OOP magnetic field at 4 K. In this exemplary case, the nonreciprocity is especially pronounced for fields below 40 mT, with the inset showing a wider magnetic field range of ±150 mT. We note a change in the overall slope of the resulting graph with the kinks at approximately ±30 mT, which marks a transition between a fast and slow decay of $I_C$ values. Figure 4(b) presents the positive critical currents of the heterostructure for different temperatures from 2 K to 5 K. With increasing temperature, we observe a decrease in the field range where the slope increases. In this figure, the corresponding negative critical currents are not depicted to distinguish the plotted curves. It is also observed here that the maximum magnitude of the critical currents at each temperature decreased from 1.25 mA at 2 K to 0.8 mA at 5 K. We note that our maximum rectification ratio ($Q \equiv 2(I_C^+ - |I_C^-|)/(I_C^+ + |I_C^-|)$) was ~5% observed at 2K. In addition, we have also observed that the Q-factor shows oscillatory behavior at some data points, which can be attributed to CGT magnetic domain reorientation or phase transitions of the finite-momentum pairing states[47]. Figure 4(c) shows the magnetization measurement of our CGT crystal for magnetic fields applied parallel to the ab-plane (black curve) and c-axis (red curve). The magnetization measurement along the c-axis saturates at a lower field, suggesting an easy axis along this crystallographic direction.

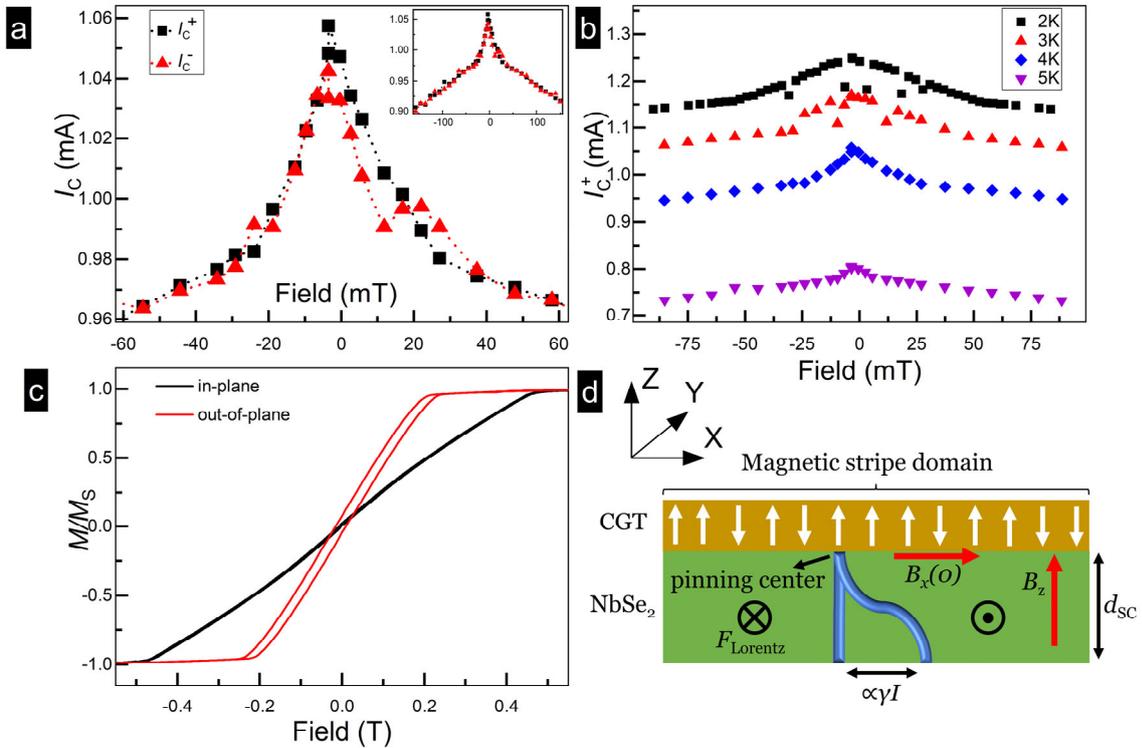

*Figure 4. a) Positive (black squares) and negative (red triangles) critical currents as a function of the field at 4 K. The nonreciprocity of these two curves is more pronounced in the range of the field from just below 0 up to nearly*

40 mT. The inset shows the critical currents for a wider magnetic field range with a noticeable slope change of the curves near ±30 mT. b) Positive critical currents as a function of the field for the temperature range 2-5 K. As expected, the critical currents decrease with increasing temperature, as well as the field range of the noticeable slope change for each temperature. c) Magnetization versus applied in-plane (black curve) and out-of-plane (red curve) magnetic fields for our CGT crystal sample at 20 K (normalized by the saturation magnetization values). d) Illustration of the flux-pinning-induced critical current anisotropy in a magnetically coupled superconductor with asymmetric interfaces when a current flows along the x-direction. Here, the magnetic stripe domains, which are oriented along the y-direction, provide the in-plane flux $B_x(0)$ while an external field $B_z$ is applied.

When $NbSe_2$ is magnetically coupled to a CGT flake, dipolar fields from the CGT magnetic texture (B <50 mT) can control the formation and dynamics of Abrikosov vortices in $NbSe_2$. Due to the asymmetric boundaries of $NbSe_2$, which is interfaced with CGT from one surface but with $SiO_2$ at the other surface, the forces at the vortex ends are not equal. This may cause a tilted core with stronger dynamics at the top $NbSe_2$/CGT interface than at the bottom $NbSe_2$/$SiO_2$. This asymmetry can be considered in explaining the nonreciprocal current-voltage characteristics[4]. As shown in Figure 4(a), we have observed a similar asymmetry between the two curves, likely due to the CGT magnetic domains that have not yet fully aligned with the applied out-of-plane field. To shed more light on this pinning mechanism, we consider a simple bilayer structure (Fig. 4(d)) and estimate the critical-current asymmetry induced by magnetic domains. In the asymmetric magnetic environment, vortices are tilted, and their length can be estimated as $L = d_{SC}\sqrt{1 + B_x^2(0)/B_z^2}$, where $d_{SC}$ is the thickness of the superconducting $NbSe_2$; $B_z$ is the external magnetic field; and $B_x(0)$ is the field component due to the asymmetric magnetic environment (dipolar field), which depends on the sweep direction of $B_z$ and applied bias current $I$. The Lorentz force is proportional to the applied current $I$ and $d_{SC}$ and can therefore be given as $F_L = \alpha d_{SC} I$. However, the pinning force is proportional to the number of pinning centers that can trap the vortex line. If the pinning centers are homogeneously distributed in the sample (no asymmetry) with an average distance of $a$ between them, then the pinning force is $F_p = \beta L/a$. In $F_L$ and $F_p$, α and β are the magnetic-field independent constants. In the presence of a bias current, the magnetic-environment field component $B_x$ changes proportionally to the current as $B_x = B_x(0) + \gamma I$, where γ is a material-specific constant. By estimating the critical current $I_c$ through the balance of the pinning force and Lorentz force at $I = I_c$, we obtain the equation $\alpha d_{SC} I_c = (\beta/a)d_{SC}\sqrt{1 + (B_x(0) + \gamma I_c)^2/B_z^2}$, whose solution is not symmetric concerning current inversion ($I_c^+ \neq I_c^-$). This simple explanation captures the observed asymmetry in an asymmetric magnetic environment $B_x(0) \neq 0$ even if pinning centers are homogeneously distributed. To obtain a deeper phenomenological understanding of the SDE, combining the transport measurements (as in this work) with direct microscopic imaging of vortex dynamics will be useful to reveal the actual operation mechanism at play[48]. We also anticipate that employing additional constrictions of varying dimensions to exploit extrinsic pinning, as in Ref. 2, will offer pathways to further increase the Q-factors.

In conclusion, we have reported the observation of the superconducting diode effect in a $NbSe_2$/$CrGeTe_3$ vdWH, where the size and asymmetry of the critical currents are modulated by the applied magnetic field, the magnetic domain texture of the CGT and inherent flux pinning in the $NbSe_2$/$CrGeTe_3$. Furthermore, this pinning effect, due to the asymmetric magnetic environments and magnetic domain transformations, can provide a means to tune and utilize the superconducting diode effect for energy harvesting from ambient radiation. More studies are required to improve the rectification efficiency and relation between intrinsic and extrinsic pinning effects using nanofabricated $NbSe_2$/$CrGeTe_3$ devices with constrictions, for instance, by modifying the shape anisotropy of CGT as well as the critical currents of $NbSe_2$. It is also interesting to look at the effects of the CGT thickness on the rectification behavior since ultrathin CGT flakes (< 10 nm) exhibit very low demagnetization fields. Our results lay the groundwork for exploring the possibility of the field-programmable superconducting diode effect and its manipulation through magnetic textures in two-dimensional systems.

## Acknowledgments

We thank Akashdeep Kamra for the fruitful discussion, the Loughborough University School of Science for financial support. We wish to acknowledge the support of the Henry Royce Institute for advanced materials for A.M. through the Student Equipment Access Scheme enabling access to the Cleanroom Facilities at University of Leeds; EPSRC Grant Number EP/R00661X/1. The work at the University of Warwick was supported by EPSRC, UK, through grants EP/T005963/1 and EP/N032128/1.

(2018).